\documentclass[prl,aps,twocolumn,superscriptaddress,fleqn]{revtex4-2}

\usepackage{
	graphicx,
	subfigure,
	placeins
}
\graphicspath{{Figures/}}

\usepackage{graphicx,subfigure,placeins,wrapfig}

\usepackage[utf8]{inputenc}
\usepackage[T1]{fontenc}
\usepackage{xcolor,booktabs,color,lipsum,makecell,microtype,fix-cm}
\usepackage[normalem]{ulem}
\usepackage{bm, amsmath,float,amssymb,upgreek,physics,braket,mhchem}
\usepackage[exponent-product=\cdot]{siunitx}

\usepackage{nicefrac} 			

\usepackage{tikz}

\usepackage{}

\usepackage{chngcntr,printlen}
\usepackage[unicode=true,colorlinks=true,citecolor=blue,urlcolor=blue]{hyperref}

\begin{document}

\title{Terahertz ratchet in graphene 2D metamaterial \\formed by  a patterned gate with an antidot array}

	\author{I. Yahniuk}
	\affiliation{Terahertz Center, University of Regensburg, 93040 Regensburg, Germany}
	
	\author{M. Hild}
	\affiliation{Terahertz Center, University of Regensburg, 93040 Regensburg, Germany}
	
	\author{L. E. Golub}
	\affiliation{Terahertz Center, University of Regensburg, 93040 Regensburg, Germany}
		
	\author{J. Amann}
	\affiliation{Terahertz Center, University of Regensburg, 93040 Regensburg, Germany}

	\author{J. Eroms}
	\affiliation{Terahertz Center, University of Regensburg, 93040 Regensburg, Germany}

	\author{D. Weiss}
\affiliation{Terahertz Center, University of Regensburg, 93040 Regensburg, Germany}

\author{Wun-Hao Kang}
\affiliation{Department of Physics and Center for Quantum Frontiers of Research and Technology (QFort), National Cheng Kung University, Tainan 70101, Taiwan}

\author{Ming-Hao Liu}
\affiliation{Department of Physics and Center for Quantum Frontiers of Research and Technology (QFort), National Cheng Kung University, Tainan 70101, Taiwan}

	\author{K. Watanabe}
\affiliation{Research Center for Electronic and Optical Materials, National Institute for Materials Science, 1-1 Namiki, Tsukuba 305-0044, Japan}

	\author{T. Taniguchi}
\affiliation{Research Center for Materials Nanoarchitectonics,  National Institute for Materials Science, 1-1 Namiki, Tsukuba 305-0044, Japan }

\author{S. D. Ganichev}
\affiliation{Terahertz Center, University of Regensburg, 93040 Regensburg, Germany}

\begin{abstract}
We report the observation of the terahertz-induced ratchet effect in graphene-based two-dimensional (2D) metamaterials. The metamaterial consists of a graphite  gate patterned with an array of triangular antidots placed under a graphene monolayer. We show that the ratchet current appears due to the noncentrosymmetry of the periodic structure unit cell. The ratchet current is generated owing to the combined action of a spatially periodic in-plane electrostatic potential and a periodically modulated radiation electric field caused by near-field  diffraction. The magnitude and direction of the ratchet current are shown to be controlled by voltages applied to both back and patterned gates, which change the lateral asymmetry, carrier type and density. The phenomenological and microscopic theories of ratchet effects in graphene-based 2D metamaterials are developed.  The experimental data are discussed in the light of the theory based on the solution of the Boltzmann kinetic equation and the calculated electrostatic potential profile.  The theory describes well all the experimental results and shows that the observed ratchet current consists of the Seebeck thermoratchet contribution as well as the linear contribution, which is sensitive to the orientation of the radiation electric field vector with respect to the triangles.  
\end{abstract}

\maketitle

\section{Introduction}
\label{introduction}

Recently, DC currents excited by terahertz (THz) radiation in two-dimensional systems with lateral 1D gratings whose period is much smaller than the radiation wavelength have attracted increasing attention~\cite{Olbrich2009ratchet,Olbrich2011,Kiselev2011,Nalitov2012, Otsuji2013,Drexler2013,Kurita2014,Budkin2014, Faltermeier2015,Olbrich2016,Popov2016,Fateev2019,Hubmann2020,BoubangaTombet2020,DelgadoNotario2020,Moench2022, Yahniuk2022,Tamura2022,Moench2023b}. These studies explored a number of fascinating effects and paved the way for the development of fast room-temperature THz detectors~\cite{Otsuji2013,Kurita2014,Faltermeier2015,Fateev2019,DelgadoNotario2020,Tamura2022,Moench2023}. A particularly promising system discussed so far is the comb-like dual-grating-gate (DGG) graphene-based field-effect transistor structures~\cite{Olbrich2011,Nalitov2012, Otsuji2013,Drexler2013,Kurita2014,Budkin2014, Faltermeier2015,Olbrich2016,Popov2016,Fateev2019,Hubmann2020,BoubangaTombet2020,DelgadoNotario2020,Moench2022, Yahniuk2022,Tamura2022,Moench2023b}. In general, the generation of a DC current in  lateral 1D periodic systems due to the action of an AC electric field requires two ingredients, namely an asymmetry of the periodic potential $V(x)$ acting on the electrons and an asymmetric amplitude of the near-field $E_0(x)$, where  $x$ is the modulation direction. The net DC electric current  is proportional to the asymmetry parameter 	$\Xi_{1\rm D} = \overline{E_0^2(x) {\mathrm{d}V(x)/ \mathrm{d}x}}$, where the overline stands for the average over the ratchet period~\cite{Olbrich2011,Nalitov2012}. The conversion of the alternating electric field into direct current in the periodically modulated structures with broken spatial symmetry belongs to the class of ratchet effects, which are relevant in different fields of physics, chemistry and biology, for reviews see, e.g.,~\cite{Juelicher1997,Linke2002,Reimann2002,Haenggi2009,Denisov2014, Ivchenko2011,Bercioux2015,Budkin2016a,Cuberto2016,Reichhardt2017,Lau2020}.  Experiments carried out in DGG structures formed by periodically repeated metal stripes of two widths deposited on various low-dimensional electronic systems, such as semiconductor quantum wells or graphene, have revealed several microscopic mechanisms of the rectified current, namely electronic ratchet effects~\cite{Olbrich2009ratchet,Ivchenko2011, Ermann2011, Bisotto2011,Olbrich2011, Kannan2012,Olbrich2016,Hubmann2020,Moench2022,Moench2023},  plasmonic ratchet/drag effects~\cite{Coquillat2010,Knap2010,Popov2011,Watanabe2013,Popov2013,Otsuji2013, Watanabe2013b,Otsuji2013b,Kurita2014,BoubangaTombet2014,Faltermeier2015,Popov2015,Rozhansky2015,Spisser2015,Olbrich2016,Koseki2016,Fateev2017b, Rupper2018,Yu2018,DelgadoNotario2020,Sai2021,Otsuji2022,Moench2022,DelgadoNotario2022,Tamura2022,Moench2023}, and the photothermoelectric effect~\cite{Tamura2022,Xu2009,Mueller2010,Yan2012,Echtermeyer2014,Cai2014}. The ratchet effect has also been demonstrated for a periodic semicircular antidot array in  Si/SiGe and AlGaAs/GaAs heterostructures excited by microwave radiation at frequencies of tens of GHz~\cite{Sassine2008,Kannan2011} and has been proposed for energy-free microwave-based signal communication~\cite{Kannan2012}. 

Here, we demonstrate that the ratchet effect can also be effectively generated by applying THz radiation to two-dimensional (2D) metamaterials formed by a patterned gate with an array of triangular shaped antidots placed under a graphene layer and having a period $d \ll \lambda$, where $\lambda$ is the radiation wavelength. Microscopically, electrons feel the low symmetry of the system through the inhomogeneous near-field formed by radiation diffraction from the boundaries of the triangular antidots.  The radiation electric field due to near-field  diffraction  $E_0(\bm r)$ is periodically modulated in 2D space with period $d$. Another field  acting on the electrons is due to the periodic asymmetric potential $V(\bm r)$ with  barriers at the boundaries of the antidots. The 2D modulation of both fields leads to the  lateral asymmetry vector parameter given by
\begin{equation}
	\label{Xi}
	\bm \Xi_{2 \rm D} =\overline{E_0^2(\bm r) \bm \nabla  {V}(\bm r)},
\end{equation}
where the overline denotes averaging over the 2D period. Although both the near-field intensity and the periodic potential are zero on average, the lateral asymmetry parameter is finite due to the low symmetry of the structure. It is clear that  $\Xi_{2 \rm D}$ vanishes  if both  profiles of $V(\bm r)$ and $E_0^2(\bm r)$ have identical coordinate dependence. However, the near-field profile is different from that of the triangles, and therefore the parameter $\bm \Xi_{2 \rm D}$ becomes nonzero. Applying  linearly polarized THz radiation to unbiased structures, we obtained DC electric currents consisting of several contributions including the polarization-independent Seebeck thermoratchet current and the linear ratchet current, which is sensitive to the orientation of the driving radiation electric field  with respect to the triangles. The asymmetric electrostatic potential in such structures can be efficiently varied by gate voltages applied to the patterned gate and/or by the uniform back gate voltages, allowing us to change the magnitude and the sign of the ratchet current in a controllable manner. The results are analyzed in terms of the phenomenological and microscopic theory of ratchet effects in periodic structures solving the Boltzmann kinetic equation. The experimental data and the theoretical results are discussed taking into account the calculated potential profile obtained from the electrostatic simulation using \textsc{FEniCS}~\cite{logg_automated_2012} based on the finite element method.

\begin{figure}[h] 
	\centering
	\includegraphics[width=\linewidth]{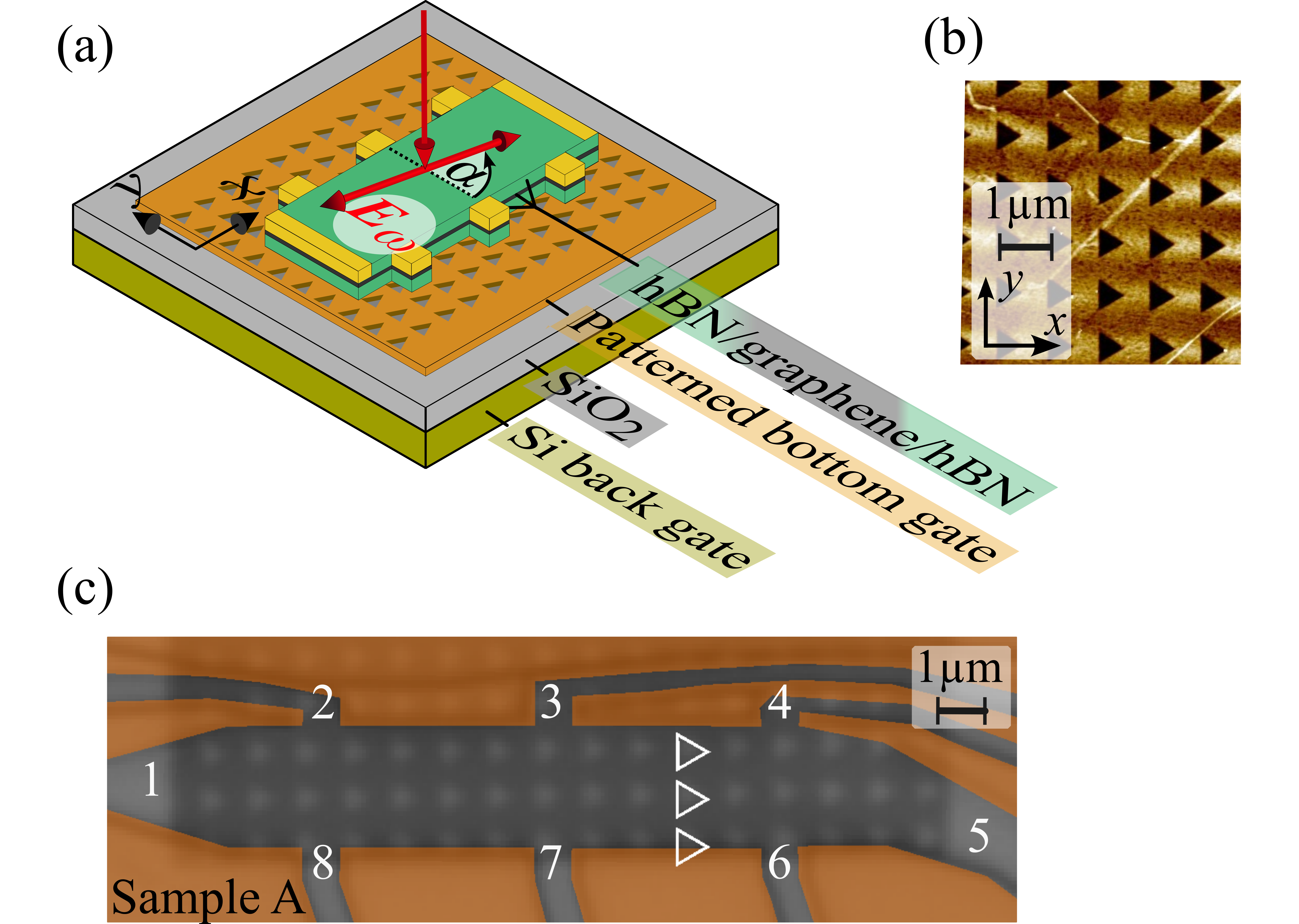}	
	\caption{Panel (a): Experimental set up and sketch of the sample. The red double arrow illustrates the radiation electric field vector $\bm E$ for  linearly polarized radiation rotated anticlockwise by the azimuth angle $\alpha$. Panel (b): AFM image of the patterned gate made of graphite with periodic array of triangular antidots of period  $d= \SI{1}{\um}$.  The antidots are holes in the bottom gate, which are  arranged in a squared lattice. Panel (c) shows a photograph of sample A (the orange area is the etching mask overlay). White triangles highlight the shape and position of one line of triangles across the Hall bar. The numbers enumerate the contacts.
}
	\label{fig1}
\end{figure}

\begin{figure}[h] 
	\centering
	\includegraphics[width=\linewidth]{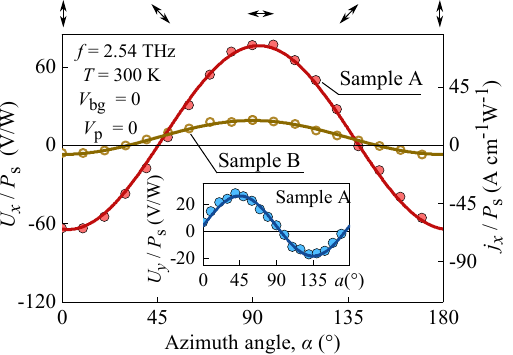}
	\caption{Dependence of the  photovoltage $U_x$ on the azimuth angle $\alpha$ measured in samples~A~and~B along the source-drain contacts 5-1. The data are normalized to power $P_s$ and obtained for zero back and patterned gate voltages. Curves are fits after Eq.\,\eqref{x-G0} with fitting parameters:   $U_{\rm L1}/P_s=\SI{70}{V\per W}$ ($U_{\rm L2} < U_{\rm L1}/10$), and $U_{\rm 0}/P_s=\SI{6}{V\per W}$  for sample~A, and  $U_{\rm L1}/P_s=\SI{13}{V\per W}$, 
	and $U_{\rm 0}/P_s=\SI{6}{V\per W}$ for sample~B. The scale on the right vertical axis corresponds to the normalized current density $j_x/P_s$ calculated for sample~A. The inset shows the polarization dependence of the normalized photovoltage $U_y/P_s$ measured in sample~A across the contacts 8-2. The gate voltages sequence is the same as for the main plot: $V_{\rm bg}=0$ and $V_{\rm p}=0$. The curve is fit after Eq.\,\eqref{y-G0} with fitting parameters 
	 $U_{\rm L2}/P_s=\SI{-22}{V\per W}$ ($U_{\rm 0} < U_{\rm L2}/6$). Arrows on top illustrate the orientation of the radiation electric field vector for several values of $\alpha$. Note that polarization dependencies described by Eqs.\,\eqref{x-G0} and\,\eqref{y-G0} are equivalent  to the theoretical Eqs.\,\eqref{j_Cs_x} and~\eqref{j_Cs_y}.	
}
	\label{fig2}
\end{figure}

\begin{figure*}[t] 
	\centering
	\includegraphics[width=\linewidth]{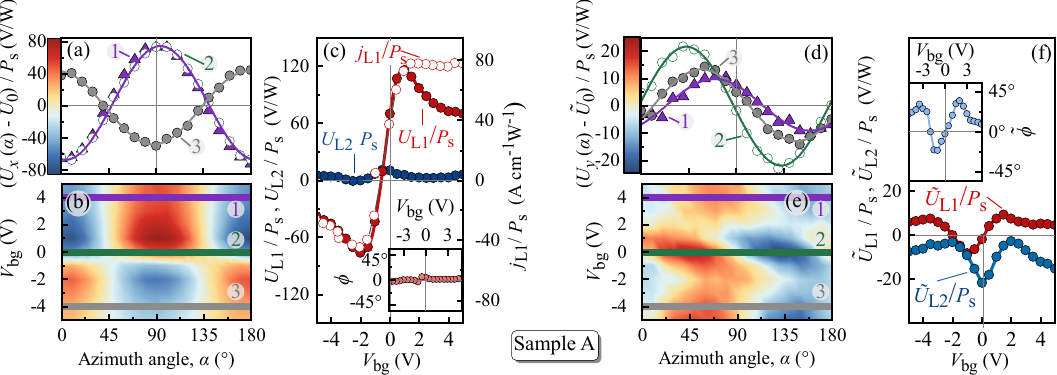}
	\caption{ Polarization dependencies of the normalized photosignal measured in sample~A along, $(U_x(\alpha) - U_0)/P_s$ (contacts 5-1), and across, $(U_y(\alpha) - U_0)/P_s$ (contacts 8-2), the Hall bar for various back gate voltages $V_{\rm bg}$ and $V_{\rm p}=0$. Note that  the polarization-independent offset $U_0/P_s$ is subtracted for visibility. The amplitudes of the signal in panel (b) are indicated by the color bar combined with the left vertical axis of  panel (a). Curves in  panel (a) are fits after Eq.\,\eqref{x-G0c1}.  Panel (c): Back gate dependencies of the fitting parameters $U_{\rm L1}/P_s$ and $U_{\rm L2}/P_s$. Lines are guide for eyes. Open red circles show the corresponding values of the normalized contributions to the ratchet current (right vertical axis).  Inset shows the gate dependence of the phase $\phi$, see Eq.\,\eqref{phi}. Panels (d) and (e): dependence of the normalized signal $(U_y(\alpha) - \tilde{U}_0)/P_s$ on the azimuth angle $\alpha$. The amplitudes of the signal in panel (e) are indicated by the color bar combined with the left vertical axis  of  panel (d). Curves in  panel (d) are fits after Eq.\,\eqref{y-G0c1}.  Panel (f): Back gate dependence of the fitting parameters $\tilde{U}_{\rm L1}/P_s$ and $\tilde{U}_{\rm L2}/P_s$. Lines are guide for eyes. Inset shows the gate dependence of the phase $\tilde{\phi}$, see Eq.\,\eqref{phitilde}. 
}
	\label{fig3}
\end{figure*}

\section{Sample and methods}
\label{samples_methods}

We fabricated our samples on a strongly $p$-doped silicon chip, which operates as a conventional global back gate. The Si back gate is separated from the rest of the sample  by a \SI{285}{nm} thick silicon dioxide, as shown in Fig.\,\ref{fig1}(a).  On top of the insulator, we deposited the key element of the structure: a patterned  bottom gate made of five layers graphene, which has  equilateral triangular antidots with a side length of 0.6\,$\upmu$m arranged in a square lattice with a period of 1\,$\upmu$m, see Fig.\,\ref{fig1}(b). The lattice was fabricated by  electron beam lithography and reactive ion etching using oxygen plasma. Monolayer graphene encapsulated in two layers of hexagonal boron nitride (hBN) was placed  on top of the lattice. The lower and upper hBN layers in sample~A (B) had  thicknesses of about \SI{40}{nm} (\SI{35}{nm}) and \SI{35}{nm} (\SI{30}{nm}), respectively. Using reactive ion etching, the samples were shaped into a Hall bar configuration, which for sample~A (B) is   \SI{16}{\micro m} (\SI{7.5}{\micro m }) long and  \SI{2.5}{\micro m} (\SI{3.5}{\micro m}) wide, and were provided with ohmic  chromium-gold contacts, see Fig.\,\ref{fig1}(a) and (c). The Hall resistance $R_{xy}$ was measured as a function of the patterned gate $V_\mathrm{p}$, ranging from \SI{-0.75}{V} to  \SI{0.75}{V} at fixed magnetic fields ($\pm 0.25$\,T), not shown. According to  $n,p = 1/(e R_\mathrm{H})$, the maximum charge densities are $4 \times 10^{11}$\,cm$^{-2}$ for electrons and  $5.5\times 10^{11}$\,cm$^{-2}$ for holes, for further details, see Appendix\,\ref{App:A}. The maximum carrier mobility for electrons and holes are $10^4$\,cm$^2$/Vs and  $0.7 \times 10^4$\,cm$^2$/Vs, respectively.

Figure~\ref{fig1}(a) sketches the experimental setup. The ratchet current was excited applying normally incident linearly polarized THz radiation from an optically pumped far-infrared molecular gas laser operating at frequency $f = \SI{2.54}{THz}$ (corresponding wavelength $\lambda = \SI{118.8}{\micro m}$ and photon energy  $\hbar\omega = \SI{10.5}{meV}$). The radiation was modulated by an optical chopper with frequency 130\,Hz. The radiation  power at the sample was about \SI{30}{mW}. The  beam spot profile  was measured by a pyroelectric camera~\cite{Ganichev1999} and had a Gaussian-like shape with a  full width at half maximum of $\SI{1.8}{mm}$. Consequently, the sample was exposed to the power $P_\mathrm{s}= P \times A_\mathrm{s}/A_\mathrm{beam}$, where $A_\mathrm{s}$ and $A_\mathrm{beam}$ are the areas of the sample and the  laser beam, respectively. To extract and study different roots of the ratchet current, the radiation electric field vector $\bm{E}$ was rotated in the sample plane using lambda-half quartz wave plates. The orientation of the vector $\bm{E}$ was defined by the azimuth angle $\alpha$ counted from the $y$-axis, see Fig.\,\ref{fig1}(a).

The samples were placed in a vacuum chamber with a $z$-cut crystal quartz window. To avoid illumination of the sample by visible and near-infrared light, the windows were covered by a black polyethylene film. All experiments were performed at room temperature. The generated photovoltage $U$ in the unbiased sample was amplified by a factor 100 and measured with standard lock-in technique. The voltage $U_x$ has been picked up from the contact pair 5-1 and $U_y$ has been picked up from contacts 8-2, with contacts 1 and 2 being grounded, respectively. The generated ratchet current density was calculated as $j=U/(R_\mathrm{s} w)$, where  $R_\mathrm{s} \ll R_{\rm in}$ is the two-point sample resistance, $R_{\rm in}$ is the input impedance of the amplifier, and $w$ is the Hall bar width.

\section{Results}
\label{results}

By irradiating the structures with the THz beam, we observed a polarization-dependent photocurrent. Figure~\ref{fig2} shows the  photocurrent density normalized to the radiation power, $j/P_s$, and the corresponding photovoltage $U \propto j$,  as a function of the angle\,$\alpha$, assigning the direction of the radiation electric field vector with respect to the height of the triangular antidots, see   Fig.\,\ref{fig1}(a). The data are obtained in two graphene structures for zero back and patterned gate voltages in the directions along the triangle height ($j_x$) and base ($j_y$). For this particular sequence of gate voltages, the polarization dependence of the signal is given by
\begin{align} 
	j_x\propto U_x &=- U_{\rm L1}\,\cos2\alpha +U_{\rm 0}  \,,\label{x-G0} \\
	j_y\propto U_y& =-\tilde{U}_{\rm L2}\,\sin2\alpha \,. \label{y-G0}
\end{align}
The application of the gate voltages changes the $\alpha$-dependence resulting in additional polarization-dependent and polarization-independent contributions. This is exemplarily shown in Figs.\,\ref{fig3}(a)--(f)  for sample~A at zero patterned gate voltage and various values of the back gate voltage. These plots reveal that the signals in $x$- and $y$-directions follow
\begin{align} 
U_x &=- U_{\rm L1}\,\cos2\alpha -U_{\rm L2}\sin2\alpha + U_{\rm 0}  \,,\label{x-G0c1} \\
U_y& =-\tilde{U}_{\rm L1}\,\cos2\alpha -\tilde{U}_{\rm L2}\sin2\alpha + \tilde{U}_{\rm 0}, \label{y-G0c1}
\end{align}
with coefficients $U_{\rm L1},\tilde{U}_{\rm L1},U_{\rm L2},\tilde{U}_{\rm L2},U_{\rm 0},\tilde{U}_{\rm 0}$ being dependent on the gate voltages $V_{\rm p}$ and $V_{\rm bg}$. In particular, the polarization-dependent photovoltage detected in the $y$-direction is formed by the interplay of two contributions having comparable amplitudes, $\tilde{U}_{\rm L1}$ and $\tilde{U}_{\rm L2}$ being characterized by different dependencies on the back gate voltage, see Fig.\,\ref{fig3}(f). This interplay leads to the gate-dependent phase shift in the $\alpha$-dependencies of $U_y$ described by 
\begin{equation}
 \tilde{U}_{\rm L1}\,\cos2\alpha +\tilde{U}_{\rm L2}\,\sin2\alpha = \sqrt{\tilde{U}_{\rm L1}^2 +\tilde{U}_{\rm L2}^2} \,\sin{2(\alpha-\tilde{\phi)}}, 
 \label{phitilde}
\end{equation}
see Figs.\,\ref{fig3}(d)--(f). The back gate dependence of the phase shift $\tilde{\phi}$ is shown in the inset in Fig.\,\ref{fig3}(f). The polarization dependence of the photocurrent detected in $x$-direction is less affected by the gate voltages. Figures~\ref{fig3}(a) and (b) show that for most back gate voltages the $U_{\rm L1}$ photovoltage contribution is substantially higher than the $U_{\rm L2}$ one [see Fig.\,\ref{fig3}(c)], and the phase shift $\phi$ in the polarization dependence  defined as 
\begin{equation}
U_{\rm L1}\,\cos2\alpha +U_{\rm L2}\sin2\alpha = \sqrt{U_{\rm L1}^2 +U_{\rm L2}^2} \cos{2(\alpha-\phi)}
\label{phi}
\end{equation}
remains close to zero, see Fig.\,\ref{fig3}(b) and  inset in Fig.\,\ref{fig3}(c). Note that a substantial deviation of the phase $\phi$ from zero is detected at high negative $V_\mathrm{bg}$ and simultaneously high positive $V_\mathrm{p}$ only, see Fig.\,\ref{fig6} (a).

In the following, we focus on the data obtained in sample~A for the $x$-direction. Since $U_{\rm L1} \gg U_{\rm L2}$, we present below a functional behavior of the $U_{\rm L1}$ and $U_0$ amplitudes only. Note that some data for the $y$-direction and for sample~B are given in Appendix~\ref{App:B}.

\begin{figure}
	\centering
	\includegraphics[width=\linewidth]{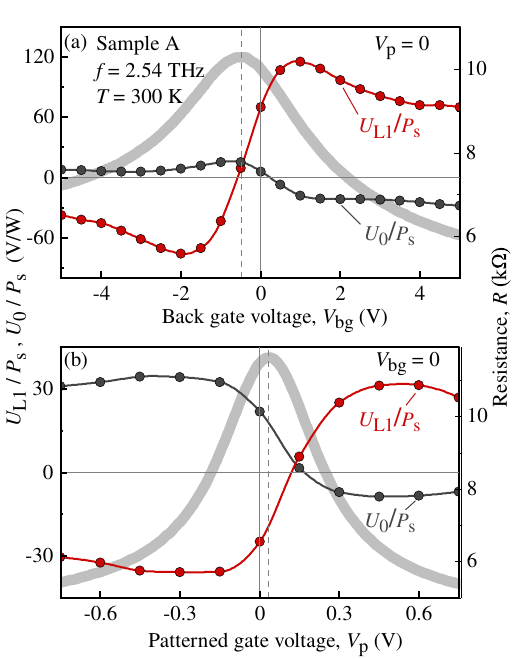}
	\caption{Panels (a) and (b): Normalized photovoltages $U_{\rm L1}/P_s$ and $U_{\rm 0}/P_s$ as a function of the back and patterned gate voltages, respectively. The data are obtained by varying one gate voltage and holding the other at zero. Gray curves show the corresponding gate voltage dependence of the two-point sample resistance $R$. }
	\label{fig4}
\end{figure}

\begin{figure}[t] 
	\centering
	\includegraphics[width=\linewidth]{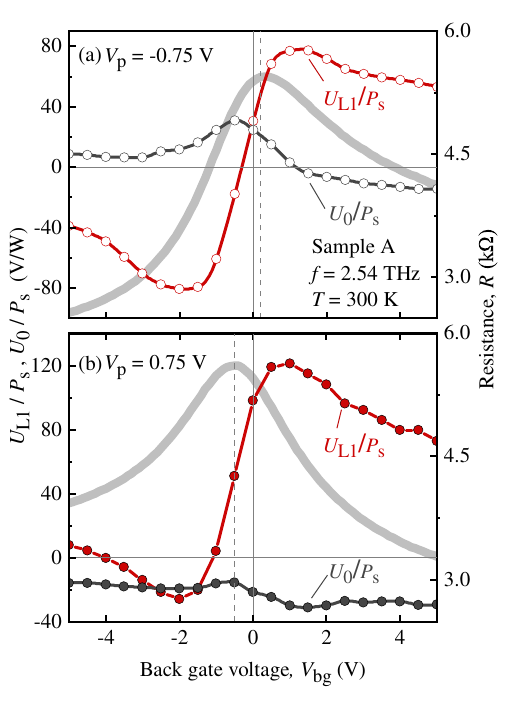}
	\caption{Normalized photovoltages $U_{\rm L1}/P_s$ and $U_{\rm 0}/P_s$ as a function of the back gate voltage. The data are obtained for $V_\mathrm{p}=\SI{-0.75}{V}$  and   $V_\mathrm{p}=\SI{0.75}{V}$, see panels (a) and (b), respectively. Gray curves show the corresponding gate voltage dependence of the two-point sample resistance $R$.
	}
	\label{fig5}
\end{figure}

\begin{figure}[t] 
	\centering
	\includegraphics[width=\linewidth]{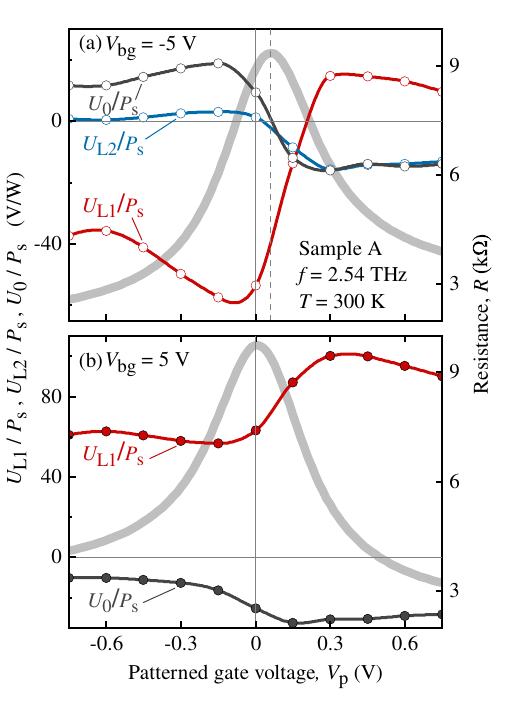}
	\caption{Normalized photovoltages $U_{\rm L1}/P_s$, $U_{\rm L2}/P_s,$ and $U_{\rm 0}/P_s$ as a function of the patterned gate voltage. The data are obtained for $V_\mathrm{bg}=\SI{-5}{V}$ (a) and   $V_\mathrm{bg}=\SI{5}{V}$; see panels (a) and (b), respectively. Gray curves show the corresponding gate dependence of the two-point sample resistance $R$.
	}
	\label{fig6}
\end{figure}

As addressed in the introduction, ratchet effects are highly sensitive to the degree of lateral asymmetry described by the parameter $\bm \Xi_{\rm 2D}$ [see Eq.\,\eqref{Xi}],  which is primarily caused by the asymmetry of the electrostatic potential. Consequently, back and patterned gate voltages should substantially change the ratchet current. Figures~\ref{fig4}--\ref{fig6} show the variation of the  $U_{\rm L1}$ and $U_{\rm 0}$ photovoltages as  the back gate and/or the patterned gate voltage is varied. All plots were obtained by holding one of the gate voltages constant and varying the other. 

Figure~\ref{fig4} shows the back gate [panel (a)] and patterned gate [panel (b)] dependencies obtained for zero $V_{\rm p}$ and $V_{\rm bg}$, respectively.  The figures reveal that the polarization-dependent and polarization-independent contributions behave similarly upon variation of both gate voltages: they change their sign near the resistance maximum, and  they decrease only slightly with $|V_{\rm bg}|$  or $|V_{\rm p}|$ at high gate voltages. Note that at high gate voltage amplitudes the sample resistance decreases, and consequently the corresponding ratchet currents, calculated as $U_x/R_s$, increase slightly with increasing $|V_{\rm bg}|$  or $|V_{\rm p}|$. Figure~\ref{fig4} also demonstrates that for most gate voltage sequences the $U_{\rm L1}$ and $U_{\rm 0}$ contributions have opposite signs.  Note that the contribution $U_{\rm L2}$  is less than 10\,\% of $U_\mathrm{L1}$ for all gate voltages except of those at which $U_{\rm L1} \approx 0$. Therefore $U_{\rm L2}$ is not shown in the plot.

Figure~\ref{fig5}(a) and (b) show the back gate voltage dependence of $U_x/P_s$ obtained for fixed positive ($V_{\rm p} = 0.75$\,V) and negative  ($V_{\rm p} = -0.75$\,V) voltages applied to the patterned gate. It can be seen that the change of the patterned gate voltage polarity does not significantly modify  the photosignal: in both plots $U_{\rm L1}$ dominates the photoresponse, reverses its sign in the vicinity of the resistance maximum and depends slightly on  $|V_{\rm bg}|$  at high back gate voltages. Note that for  positive patterned gate voltages, in contrast, at high negative back gate voltages the amplitude $U_{\rm L1}$ decreases and  for $V_{\rm p} = 0.75$\,V even changes its sign at $V_{\rm bg} \simeq -4$\,V. For $V_{\rm p} = 0.75$\,V the polarization-independent photoresponse $U_{\rm 0}$ is negative for all back gate voltages.

Figures~\ref{fig6}(a) and~(b) present the patterned gate voltage dependence obtained for large back gate voltages with opposite polarities of $-5$\,V and 5\,V, respectively. Also under these conditions, the $U_{\rm L1}$ contribution dominates the photoresponse for nearly all gate voltages except for $V_{\rm p} > 0.2$\,V at which all three contributions  $U_{\rm L1}$,  $U_{\rm L2}$, and  $U_{\rm 0}$ have comparable amplitudes, see Fig.\,\ref{fig6}(a).  While at negative back gate voltage the signals change their sign in the vicinity of the resistance maximum, at positive $V_{\rm bg} = 5$\,V both $U_{\rm L1}$  and  $U_{\rm 0}$ keep their signs, see Fig.\,\ref{fig6}(b). At the same time, the magnitude of these contributions changes significantly in the gate voltage range where the structure resistance approaches its maximum value. Again, a reduction of the resistance increases the current magnitudes at high gate voltage amplitudes.

\section{Discussion}
\label{discussion}

To analyse the data obtained, we first present a symmetry analysis and microscopic theoretical treatment. In the presence of a substrate,  a uniform gate, or adatoms deposited on one side of the sample, the symmetry of graphene is reduced to the non-centrosymmetric group C$_{6v}$, because the equivalence of the $z$ and $-z$ directions is removed. Even for this symmetry class,  the generation of direct current under normal incidence radiation is forbidden, and indeed not observed, for a review see~\cite{Glazov2014}. In general, normal incident radiation can give rise to edge photocurrents, which can be detected by measuring the signal from the pair of contacts along the edges and have opposite directions for opposite edges~\cite{Glazov2014,Candussio2020}. However, the edge currents cannot be responsible for the discussed signals, because they disappear when the radiation electric field is oriented along or perpendicular to the edges, for which orientations, in our experiments, the signal $U_x$ approaches its maximum, see Figs.\,\ref{fig2} and\,\ref{fig3}(a)--(c). These data, together with the above arguments, provide evidence that the observed signals are caused by the presence of the periodic triangular antidots.

The system of infinite triangular antidots array arranged in a quadratic lattice has $C_s$ point symmetry with only one non-trivial symmetry element, namely a reflection in the $(zx)$ plane perpendicular to the structure plane and to the triangles' bases. At normal light incidence, the following relations between the net photocurrent density $\bm j$ and the Stokes parameters of the radiation are allowed in $C_s$ symmetry:
\begin{align}
	\label{j_Cs_x}
	&j_x = \Xi_x \qty[\chi_0 
	+ \chi_1 \qty(\abs{e_x}^2-\abs{e_y}^2)],
	\\
	\label{j_Cs_y}
	&j_y =\Xi_x 
	{\chi}_2 \qty(e_xe_y^*+e_x^*e_y),
\end{align}
where $\bm e$ is the unit polarization vector of the incident radiation $\bm E(t)=E_0\bm e \exp(-i\omega t)+c.c.$, and $\Xi_x$ is the $x$-component of the vector $\bm \Xi_{\rm 2D}$ defined by Eq.\,\eqref{Xi}.  The value $\chi_0$ describes the polarization-independent contribution to the ratchet current,  while $\chi_1$ and ${\chi}_2$ describe the contributions depending on the linear-polarization  orientation. These dependencies describe well the experimental findings in the ungated structures ($V_{\rm p} = V_{\rm bg}=0$). Indeed, in the described experiments the polarization-dependent parts are given by $\abs{e_x}^2-\abs{e_y}^2= -\cos 2\alpha$ and $e_xe_y^*+e_x^*e_y=- \sin2\alpha$. Corresponding fitting curves are shown in Fig.\,\ref{fig2}.

Microscopically, the ratchet current is caused by the simultaneous action of the asymmetric static 2D-periodic potential $V(\bm r)$  and the  2D-periodic THz near-field with amplitude ${E}_0(\bm{r})$, formed by radiation diffraction at the edges of the triangular antidots. Although both the near-field and the periodic potential are zero on average, the  parameter $\bm \Xi_{\rm 2D}$, see Eq.\,\eqref{Xi}, is finite and captures the $C_s$ symmetry of the system\,\footnote{Note that, in the structure under study, the size of the triangles is much larger than the carrier mean free path. Therefore, the mechanism suggested in Ref.~\cite{Chepelianskii2008} and used for description of experimental data in Refs.~\cite{Sassine2008,Kannan2011} is not applicable to our system.}.

\begin{figure}[t]
	\centering
	\includegraphics[]{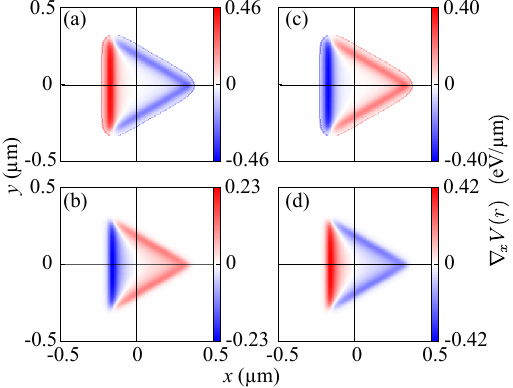}
	\caption{Partially differentiated onsite potential in one period ($\SI{1}{\um}$)  as a function of $x$ and $y$, considering (a) $(V_{\mathrm{bg}}, V_\mathrm{p}) = (-2, 0)\ \mathrm{V}$, (b) $(V_{\mathrm{bg}}, V_\mathrm{p}) = (1.5, 0)\ \mathrm{V}$, (c) $(V_{\mathrm{bg}}, V_\mathrm{p}) = (0, -0.15)\ \mathrm{V}$ and (b) $(V_{\mathrm{bg}}, V_\mathrm{p}) = (0, 0.5)\ \mathrm{V}$ relevant to experiments. }
	\label{fig7}
\end{figure}

By treating the static periodic potential and the radiation electric field as small perturbations, the microscopic description of the current formation can be reduced to the  problem of 1D modulation which was previously considered for graphene  in Refs.~\cite{Nalitov2012,Olbrich2016}. Therefore, the 2D character of the modulation in the present discussion is accounted for by the lateral asymmetry parameter $\bm \Xi_{\rm 2D}$, while the parameters $\chi_{0,1,2}$ are determined solely by the properties of the 2D carriers in graphene above the periodic gate. These parameters are obtained in the kinetic equation approach~\cite{Nalitov2012,Hubmann2020} for an arbitrary relation between the radiation frequency $\omega$ and the transport relaxation time $\tau_{\rm tr}$, see Appendix~\ref{App:C}. In the high-frequency limit $\omega\tau_{\rm tr} \gg 1$ relevant to the experimental conditions, the obtained expressions give the following results: 
\begin{align}
	&\chi_1 = \chi_2 = - {q^3 v_0^2 \tau_{\rm tr}\over 2\pi \hbar^2 \omega^2 \varepsilon_{\rm F}},\nonumber \\
	&\chi_0 = \chi_1 \qty(1 - {16\pi^2 k_{\rm B}T \tau_T\over 9\varepsilon_{\rm F}\tau_{\rm tr}}).
	\label{microscopic}
\end{align}
Here $v_0$ is the Dirac fermion velocity in graphene, $\varepsilon_{\rm F}$ is the Fermi energy, $q$ is the elementary charge, which is positive for holes and negative for electrons, $T$ is the lattice temperature, and $\tau_{T}$ is the electron temperature relaxation time related with the electron-phonon interaction~\cite{Olbrich2016}. The coefficients $\chi_{1,2}$  are obtained in the dynamic charge density redistribution (DCDR) model and describe the polarization-dependent ratchet currents, while the coefficient $\chi_0$ accounts for the polarization-independent ratchet current, see Eqs.\,\eqref{j_Cs_x} and\,\eqref{j_Cs_y}. While the first term in $\chi_0$ is also caused by the DCDR mechanism, the second term describes the so-called Seebeck ratchet effect caused by electron gas heating and simultaneous spatial charge  redistribution~\cite{Olbrich2016}. We see that $\chi_{1,2}$ depend on the carrier density $n$ via the ratio $\tau_{\rm tr}/\varepsilon_{\rm F}$. While for scattering by long-range Coulomb impurities this factor is independent of $n$, for short-range scatterers we have $\chi_{1,2} \propto 1/n$, i.e. the linear polarization driven ratchet currents decrease with increasing density. Note that $\chi_0$ has the opposite sign relative to $\chi_{1,2}$ if the Seebeck contribution dominates the polarization-independent ratchet current.

Although the parameters $\chi_{0,1,2}$ are given by Eqs.\,\eqref{microscopic}, the total ratchet current in the studied system is formed in a more complex way than in 1D-modulated systems considered in Refs.~\cite{Nalitov2012,Olbrich2016}. This is because in our 2D system, both patterned and uniform backgates form two types of regions with different carrier densities ($n$ and $n^*$ or $p$ and $p^*$) or even carrier types ($n$ and $p^*$ or $p$ and $n^*$). Consequently, the ratchet current is generated in two areas with free carriers of different microscopic properties and different asymmetry parameters,  $\bm \Xi^*_{\rm 2D}$ and  $\bm \Xi_{\rm 2D}$, for carriers in graphene above and outside triangular antidots, respectively. Therefore, the net ratchet current is a sum of two contributions which can be different in amplitude and even in sign.

While the results in ungated samples are well described by the phenomenological  Eqs.\,\eqref{j_Cs_x} and\,\eqref{j_Cs_y}, additional contributions are clearly detected  at non-zero gate voltages, see Fig.\,\ref{fig3} and Eqs.\,\eqref{x-G0c1},\,\eqref{y-G0c1}. This observation reveals the symmetry reduction from C$_s$ to $C_1$.  This is in agreement with the analysis of the AFM image of the studied structure, see Fig.\,\ref{fig1}(c), which  shows that the graphene flake covers only a part of the upper row of the triangular antidots array, which consequently removes the ($zx$) mirror reflection of the whole structure -- the last non-trivial symmetry operation. As a result of this symmetry reduction, an additional lateral asymmetry parameter $\Xi_y$ (the $y$-component of the parameter $\bm \Xi_{\rm 2D}$, see Eq.\,\eqref{Xi}) emerges. It gives rise to new contributions to the net photocurrent 
\begin{align}
	&\delta j_x = \Xi_y
	{\chi}_2 \qty(e_xe_y^*+e_x^*e_y),
	\\
	&\delta j_y =\Xi_y  \qty[\chi_0 
	- \chi_1 \qty(\abs{e_x}^2-\abs{e_y}^2)].
\end{align}
It is noteworthy that the values $\chi_0$, $\chi_1$, and ${\chi}_2$ are the same here as in Eqs.\,\eqref{j_Cs_x} and\,\eqref{j_Cs_y}, since they are determined by the properties of the graphene layer and are independent of the lateral asymmetry potential. Consequently, the total current is given by
\begin{align}
	\label{j_C1_x}
	j_x = &\Xi_x \qty(\chi_0  
	- \chi_1 \cos 2\alpha) 
	-\Xi_y
	 {\chi}_2 \sin 2\alpha,
	\\
	\label{j_C1_y}
	j_y = &-\Xi_x
	{\chi}_2 \sin 2\alpha + 
	\Xi_y  \qty(\chi_0 
		+ \chi_1 \cos 2\alpha).
\end{align}
These dependencies describe all experimental results well, see Fig.\,\ref{fig3} and Appendix~\ref{App:B}.  The fitting curves of the azimuth angle dependencies are shown in Figs.\,\ref{fig3}(a) and\,(d), where the fitting parameters are $U_{\rm L1} \propto \Xi_x \chi_1$, $U_{\rm L2} \propto \Xi_y \chi_2$, $U_0 \propto \Xi_x \chi_0$, $\tilde{U}_{\rm L1} \propto \Xi_y \chi_1$, $\tilde{U}_{\rm L2} \propto \Xi_x \chi_2$, and $\tilde{U}_0 \propto \Xi_y \chi_0$. Note that the reduction in symmetry primarily affects the $y$-component of the ratchet current (i.e. the photocurrent across the Hall bar), a direction in which the edge effects are expected to have the largest influence. Indeed, Figs.\,\ref{fig3}(d)--(f) and Figs.\,\ref{figB1}--\ref{figB3} in  Appendix~\ref{App:B} show that the photovoltages specific for  C$_1$ symmetry   are mostly pronounced when measured in the $y$-direction: the contribution $\tilde{U}_{\rm L1}\cos2\alpha$  dominates the signal even at high positive gate voltages. Furthermore, we detected  a substantial polarization-independent contribution in this direction, which is also a prerogative of C$_1$ symmetry, see Appendix~\ref{App:B}. In contrast, for the photovoltages measured in the $x$-direction, the photovoltage $U_{\rm L2}\sin2\alpha$  specific for $C_1$ symmetry   is significantly smaller than $U_{\rm L1}\cos2\alpha$ for the most gate voltage sequences. A comparable $U_{\rm L1}$ and $U_{\rm L2}$  are  detected only for gate voltage sequences where the  $U_{\rm L1}$-term changes its sign (not shown). We emphasize that, as  discussed above, the net current is formed in two regions of graphene, above and outside the triangular antidots.

Now we discuss the observed gate dependencies of the ratchet currents, see Fig.\,\ref{fig3}(c) and Figs.\,\ref{fig4}--\ref{fig6}. According to the theory developed above, the sign and magnitude of the ratchet currents are determined by the lateral asymmetry parameters $\Xi_x$ and  $\Xi_x^*$, the sign and the density of charges in the regions above ($n^*,p^*$) and outside ($n,p$) the triangles, lattice temperature, scattering times, and the radiation frequency and intensity, see Eqs.\,\eqref{microscopic}.  A crucial condition for the generation of ratchet currents is a non-zero lateral asymmetry parameter $\bm \Xi_{2 \rm D}$, and thus the presence of an asymmetric electrostatic potential. In our experiments, a  ratchet current is also present  when both gate voltages are zero, see Figs.\,\ref{fig4}(c) and\,\ref{fig5}. However, even in this case the asymmetry is created by the built-in potential caused by the conducting patterned gate deposited under the graphene.  According to the theory, see Eqs.\,\eqref{microscopic}, the signs of polarization-dependent and independent ratchet currents are opposite provided that the Seebeck contribution is dominant in the polarization-independent current.  This is observed for almost all sequences of gate voltages, see Figs.~\ref{fig4}--\ref{fig6}. 

To investigate the effect of gate voltages on the asymmetry of the electrostatic potential, we performed the simulation by using \textsc{FEniCS}~\cite{logg_automated_2012} which is based on the finite-element method. After extracting the electric displacement field near  graphene, the spatially dependent capacitance of each gate was obtained, see Appendix~\ref{App:D} for details. Figure~\ref{fig7} shows exemplarily in-plane distributions of $\nabla_x V(\bm r)$ calculated for four different sequences of patterned gate and back gate voltages relevant to the experiments. The curves show that the asymmetry along the $x$-direction is indeed present and can be reversed by reversing the polarity of one of the gates. In particular, in the discussed metamaterial the uniform bottom gate potential leads to a change of $\bm \nabla V$ and thus of the lateral asymmetry parameter. This is because the applied back gate voltage is periodically screened by the graphite gate. It can even lead to the formation of lateral $p-n$ junctions. Note that the built-in potential was not included in the simulation.

Figures~\ref{fig3}--\ref{fig6} show the following main qualitative features: i)\,a strong variation of the ratchet current amplitudes in the vicinity of the resistance maximum region; ii)\,a weak back and patterned gate dependence outside this gate voltage range; iii)\,the ratchet current may or may not change its sign upon switching the gate voltage polarity; iv)\,in most cases, the polarization-dependent and independent contributions have opposite signs for fixed gate sequences. Analogous arguments also apply to $U_y$, whose gate voltage dependencies are presented in Appendix~\ref{App:B}. All these features can be qualitatively explained by the above theory. The  sign of the ratchet current is defined by the product $q^3 \Xi_x$ and/or $q^{*3} \Xi_x^*$, see  Eqs.\,\eqref{j_Cs_x}--\eqref{microscopic}. Assuming that the spatial dependence of the near-field of diffraction does not reverse upon variation of the gate voltages, the sign of the ratchet current is solely defined by the sign of $q$ (negative for electrons and positive for holes) and $\nabla_x V$. Therefore, if they both change sign, then the ratchet's current direction remains unchanged. Otherwise, if one of the signs holds and the other reverses, the current direction is changed. As discussed above, the net current is a sum of the currents generated in two areas, above and outside the triangles. For this reason, for example, reversing the patterned gate voltages would change both, the sign of the electrostatic potential and the sign of $q$ ($p \to n$) in the region outside the triangles, leaving the current direction unchanged, see e.g. $U_{\rm L1}$ in Fig.\,\ref{fig6}(b). Note that the part of the current that changes its sign with the patterned gate voltage variation, is responsible for the observed substantial change of the ratchet current amplitudes in the vicinity of the resistance maximum region, see Fig.\,\ref{fig6}(b). At the same time,  the sign of  $\nabla_x V$ is inverted above the triangles, while the carrier type remains unaffected, causing the sign inversion of the net current, see e.g. $U_{\rm L1}$ in Fig.\,\ref{fig4}(b).

\section{Summary}
\label{summary}

In summary, we demonstrate that the ratchet current can be efficiently generated in graphene-based 2D metamaterials excited by linearly polarized terahertz radiation. The directed current in response to an alternating electric field is caused by the asymmetry of the periodic electrostatic potential and the near-field  diffraction formed by a patterned gate with an array of triangular antidots placed under a graphene layer and having a period much smaller than the radiation wavelength. Experimental data together with the developed theory demonstrate that the ratchet current consists of two contributions caused by the polarization-independent Seebeck ratchet effect and the DCDR mechanism. We show that in the studied structures with the patterned gate placed between graphene and the uniform back gate, the back gate voltage also introduces an asymmetry of the electrostatic potential. This is confirmed by the calculation using \textsc{FEniCS}~\cite{logg_automated_2012} based on the finite-element method. The presence of two asymmetric potentials, which affect the lateral asymmetry parameter and lead to a periodic modification of the carrier density (carrier types) in the areas above and outside the triangular antidots, causes a complex dependence of the ratchet current on the gate voltages. Our analysis shows that while increasing the gate voltage increases the lateral asymmetry parameter, it does not lead to a corresponding increase of the ratchet current. This is due to the linear dispersion of graphene, for which increasing the gate potentials decreases the ratchet current.

\section{Acknowledgments}
\label{acknow}

The authors thank V.V. Bel'kov for valuable discussions.  We acknowledge the financial support of the Deutsche Forschungsgemeinschaft (DFG, German Research Foundation) via Project-ID  314695032 – SFB 1277 (Subprojects  A04 and A09),  Project-ID 426094608, and of the Volkswagen Stiftung Program (97738). W.-H.K. and M.-H.L. gratefully acknowledge National Science and Technology Council of Taiwan (grant number: NSTC 112-2112-M-006-019-MY3). K.W. and T.T. acknowledge support from the Elemental Strategy Initiative conducted by the MEXT, Japan (Grant Number JPMXP0112101001) and JSPS KAKENHI (Grant Numbers 19H05790 and JP20H00354).

\appendix
\counterwithin{figure}{section}
\setcounter{figure}{0}

\section{Resistance map of sample A}
\label{App:A}

\begin{figure}[h] 
	\centering
	\includegraphics[width=\linewidth]{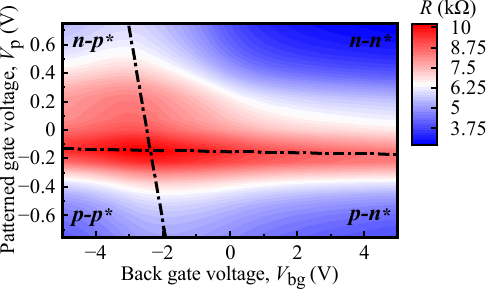}	
\caption{ Map of the two-terminal resistance $R$ (with contours and color scale) as a function of both  gate voltages $V_\mathrm{p}$ and $V_\mathrm{bg}$.   The black dashed-dot lines represent the behavior (position) of the charge neutrality point. The regions of maximum voltages applied to gates $V_{\rm p}$ and $V_{\rm bg}$ ($p-p^*$, $n-p^*$, $p-n^*$, $n-n^*$) are labeled as guides for the eyes. Here, $p$- and $n$- correspond to hole- and electron- densities in the active area of the monolayer graphene, respectively. Designations of $p^*$- and $n^*$- represent the hole-type carriers and electron-type carriers in the triangular regions.
	}
	\label{figA1}
\end{figure}

Figure~\ref{figA1} shows the two-terminal sample resistance $R$ as a function of  a patterned, $V_\mathrm{p}$, and back, $V_\mathrm{bg}$, gate voltages. The data are obtained applying 	the standard lock-in technique with low modulation frequency ($\sim\,11$\,Hz) and the source-drain current \SI{0.1}{\um}.  The two-terminal resistance is controlled by both $V_\mathrm{p}$ and $V_\mathrm{bg}$, where $V_\mathrm{bg}$ mainly governs the carrier concentration in the graphene regions above the triangular holes, while $V_\mathrm{p}$ influences the graphene regions around the holes. Therefore, by adjusting both gate voltages, the carrier density and modulation potential can be set indepedently \cite{Drienovsky2018}. Charge neutrality across the entire sample is achieved at $V_\mathrm{p}\approx -0.15\ \mathrm{V},\ V_\mathrm{bg}\approx -2.4\ \mathrm{V}$, i.e., at the crossing point of the dash-dotted lines. This point deviates slightly from the expected zero voltages due to undesired background doping, which changes upon thermal cycling. Therefore, the gate positions in the resistance map in Fig.\,\ref{figA1} cannot be directly matched to the photovoltage data, as they were obtained in different cooldowns. When one gate voltage is fixed, sweeping the other gate yields resistance traces with a resistance maximum, similar to the charge neutrality point in unmodulated graphene. Those maxima occur along the two dash-dotted lines, which divide the map into four quadrants. Two of them correspond to the same carrier type inside and outside the triangular regions, the other two designate the bipolar regime, in which the carrier type inside and outside the triangular regions are opposite. We label those quadrants using $n$ and $p$ for electron and hole type, and mark the triangular regions with an asterisk ('*'). To determine the carrier densities, we carried out  four-probe measurements of $R_{xx}$ and $R_{xy}$ for various gate voltages $V_\mathrm{p}$ and $V_\mathrm{bg}$ at magnetic field $B = \SI{0.25}{T}$. The Hall concentration was experimentally determined as $n_\mathrm{H} = 1/eR_\mathrm{H}$, where $R_\mathrm{H}$ is the Hall coefficient. In  case of a fixed zero voltage $V_\mathrm{bg}$ on the back gate electrode, the 2D carrier density is also empirically given by $n = 6.3 \times 10^{11}  (V_\mathrm{p}-V_0)$\,cm$^{-2}$, where $V_\mathrm{p}$ and a gate offset $V_0$ are taken in volts.

\section{Gate dependencies of the photovoltage $U_y$ measured across Hall bar and additional results for the second sample}
\label{App:B}

\subsection{Gate dependencies of the photovoltage $U_y$ measured across Hall bar obtained for sample A}
\label{App:B1}

\begin{figure}[t] 
	\centering
	\includegraphics[width=\linewidth]{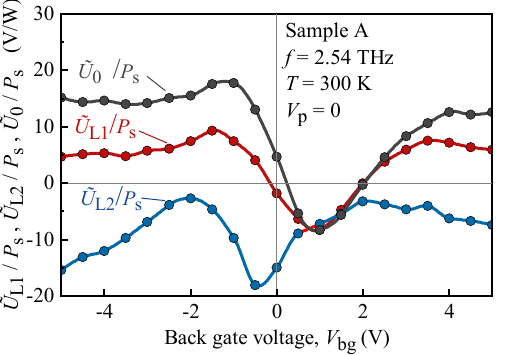}
	\caption{ Panels (a) and (b): Normalized photovoltages  $\tilde{U}_{\rm L1}/P_s$,  $\tilde{U}_{\rm L2}/P_s$ and $\tilde{U}_{\rm 0}/P_s$ as a function of the back gate voltage. The data are obtained varying one gate voltage and holding zero the other.	}
	\label{figB1}
\end{figure}

The polarization dependence of the photosignal $U_y$ measured across the contact pair 8-2 is well described by Eq.\,\eqref{y-G0c1} being in full agreement with Eq.\,\eqref{j_C1_y} of the theory developed for C$_1$ symmetry.  Figure~\ref{figB1} shows the photosignal $U_y$ measured across the contact pair 8-2 as a function of the  back gate voltage. The back gate dependencies of the individual contributions $\tilde{U}_{\rm L1}$, $\tilde{U}_{\rm L2}$ and $\tilde{U}_0$ are obtained for zero patterned gate voltage. In line with the discussion presented in Sec.\,\ref{discussion} the contribution $U_y \propto \tilde{U}_{\rm L2}$ dominates the photoresponse for zero potentials at both gates, whereas sweeping the back gate voltage results in a complex variation of the signal magnitude and sign. Figure~\ref{figB1} demonstrates that the amplitudes $\tilde{U}_{\rm L1}$ and $\tilde{U}_{0}$ twice reverse their sign while the sign of the amplitude $\tilde{U}_{\rm L2}$ remains unchanged in the whole range of the applied gate voltages.

\begin{figure}[t] 
	\centering
	\includegraphics[width=\linewidth]{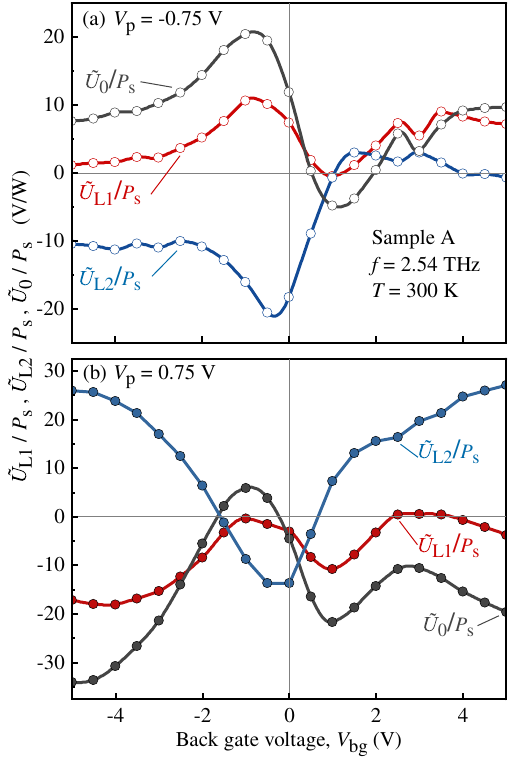}
	\caption{Normalized photovoltages  $\tilde{U}_{\rm L1}/P_s, \tilde{U}_{\rm L2}/P_s$ and $\tilde{U}_{\rm 0}/P_s$ as a function of the patterned gate voltage. The data are obtained for $V_\mathrm{p}=\SI{-0.75}{V}$ (a) and   $V_\mathrm{p}=\SI{0.75}{V}$ (b).}
	\label{figB2}
\end{figure}

\begin{figure}[t] 
	\centering
	\includegraphics[width=\linewidth]{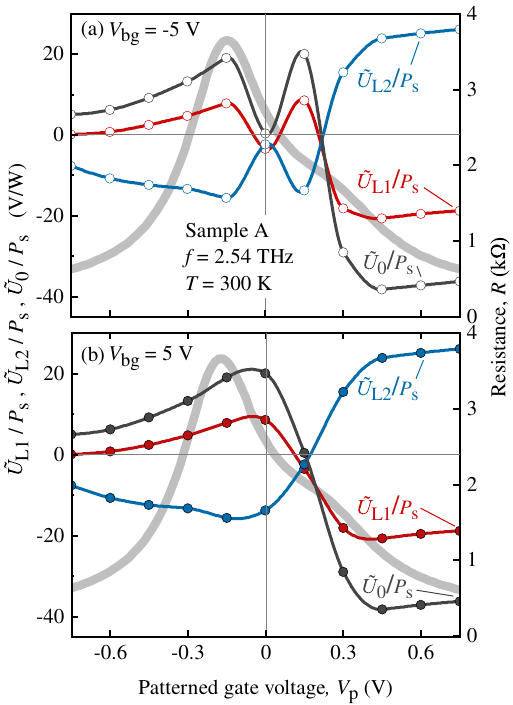}
	\caption{Normalized photovoltages  $\tilde{U}_{\rm L1}/P_s, \tilde{U}_{\rm L2}/P_s$ and $\tilde{U}_{\rm 0}/P_s$ as a function of the patterned gate voltage. The data are obtained for $V_\mathrm{bg}=\SI{-5}{V}$  and   $V_\mathrm{bg}=\SI{5}{V}$; see panels (a) and (b), respectively. 	}
	\label{figB3}
\end{figure}

Gate dependencies obtained varying one of the gate voltages and holding the other confirm a complexity of the gate dependencies. A double inversion of the signal sign is detected in the back gate dependencies for both positive and negative patterned gate voltages see Fig.\,\ref{figB2}, whereas the signal change its sign only once if a high positive back gate voltage is applied and the patterned gate voltage is varied, see Fig.\,\ref{figB3}(b). A single sign inversion of $V_y (V_{\rm p})$ is also detected for high negative back gate voltage, see Fig.\,\ref{figB3}(a). However, this dependence is characterized by  a highly nonmonotonic behavior of all contributions in the vicinity of the resistance maximum.

As discussed above, these complex behavior is caused by the interplay of the lateral asymmetry parameter $\Xi_{\rm 2D}$ and the carrier density/type in the region above triangles, which is controlled by the back gate voltage.

\subsection{Polarization and gate dependencies of the ratchet current obtained for sample B.}
\label{App:B2}

\begin{figure}[t] 
	\centering
	\includegraphics[width=\linewidth]{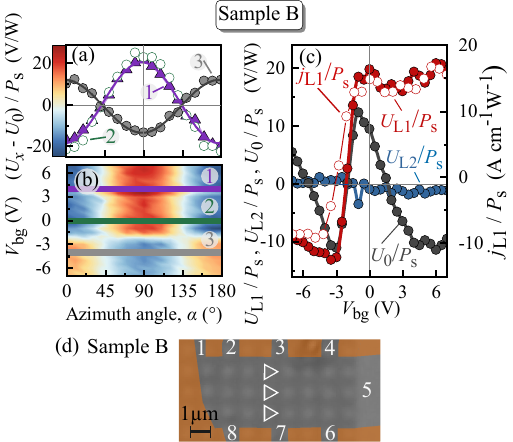}
	\caption{Panels (a) and (b) show the polarization dependencies of the normalized photosignal $(U_x(\alpha) - U_0)/P_s$ measured in sample~B along the Hall bar (contacts 5-1) for different back gate voltages $V_{\rm bg}$ and $V_{\rm p}=0$. Note that  a small polarization-independent offset $U_0/P_s$ is subtracted for visibility. The amplitudes of the signal in panel (b) are indicated by the color bar combined with the left vertical axis of  panel (a). Curves in  panel (a) are fits after Eq.\,\eqref{x-G0c1}.  Panel (c): Back gate dependencies of the fitting parameters $U_{\rm L1}/P_s$ and $U_{\rm L2}/P_s$. Lines are a guide for the eyes. Right ordinate show the corresponding values of the normalized ratchet current density $j_x/P_s$.  Panel (c) show a photograph of the samples. White triangles highlights the shape and position of one line of triangles across the Hall bar.}
	\label{figB4}
\end{figure}

Figures~\ref{figB4}(a) and~(b) show the polarization dependence of the photocurrent measured in sample~B along the source-drain pair of contacts. The data are obtained for different values of the back gate voltage holding $V_\mathrm{p}=0$. The data are well described by Eq.\,\eqref{x-G0c1} being in  full agreement with Eq.\,\eqref{j_C1_x} developed for C$_1$ symmetry.  To highlight the polarization dependence, in panels (a) and (b) we subtracted a small polarization-independent contribution.  The overall behavior of the photocurrent is similar to that detected in sample A. Figure~\ref{figB4}(c)  shows the back gate dependence of the individual contribution obtained by holding the patterned gate voltage at zero. The figure reveals that the contributions $U_{\rm L1}$ and $U_{\rm 0}$ dominate the signal in the whole range of the gate voltages. The former one reverses its sign in the vicinity of the resistance maximum and becomes almost constant at high back gate voltages. The polarization-independent contribution, by contrast, reverses sign three times.

\section{Extended description of the microscopic theory}
\label{App:C}

The microscopic theory of the ratchet effect in graphene-based systems~\cite{Nalitov2012,Hubmann2020} yields the following expressions for $\chi_{0,1,2}$:
\begin{subequations}
	\label{eq:chi}
	\begin{align}
		&\chi_0 = \chi_0^\mathrm{S} + {q^3 v_0^2 \over 2\pi \hbar^2} \qty(\text{Re}S_1-\text{Im}S_2),
		\\
		& \chi_1= \chi_2 = - {q^3 v_0^2 \over 2\pi \hbar^2} \qty(\text{Im}S_2+\delta S),
	\end{align}
\end{subequations}
where $v_0$ is the Dirac fermion velocity in graphene and $q$ is the elementary charge. The contribution $\chi_0^\mathrm{S}$ describes the so-called Seebeck ratchet effect caused by  electron gas heating and simultaneous spatial charge  redistribution and is given by~\cite{Olbrich2016}
\begin{equation}
	\chi_0^\mathrm{S} = {8\pi q^3 v_0^2 \tau_{\rm tr}^2  \tau_T k_{\rm B}T \over 9\hbar^2 \varepsilon_{\rm F}^2[1+(\omega\tau_{\rm tr})^2]},
\end{equation}
where $T$ is the lattice temperature, and $\tau_{T}$ is the electron temperature relaxation time related with the electron-phonon interaction~\cite{Olbrich2016}. The polarization-dependent contributions to the ratchet current in Eqs.\,\eqref{eq:chi} are not related to the electron gas heating, but determined by the dynamic charge-density redistribution (DCDR) controlled solely by elastic scattering processes~\cite{Faltermeier2018}. In particular, these terms are controlled by the dependence of the elastic scattering times on the electron energy, i.e. they are sensitive to the disorder potential. The values $S_{1,2}$ and $\delta S$ were calculated in Refs.~\cite{Nalitov2012} and~\cite{Hubmann2020}:
\begin{align}
	&S_1 = \varepsilon_{\rm F}^3 \qty({\tau_{\rm tr}\over \varepsilon_{\rm F}})'\tau_{2\omega}\qty({\tau_{{\rm tr}\omega}\over \varepsilon_{\rm F}})' - {1\over 2}\qty[\varepsilon_{\rm F}^2 \qty({\tau_{\rm tr}\over \varepsilon_{\rm F}})'\tau_{2\omega}\tau_{{\rm tr}\omega}]', 
	\\
	&S_2 = {\qty( \varepsilon_{\rm F} \tau_{\rm tr})'\qty(\varepsilon_{\rm F} \tau_{{\rm tr}\omega})'\over \omega\varepsilon_{\rm F}} - {\qty[\qty( \varepsilon_{\rm F} \tau_{\rm tr})' \tau_{{\rm tr}\omega}]'\over 2\omega},
	\\
	&
	\delta S = {\qty[\varepsilon_{\rm F}^3 \qty(\tau_{\rm tr}/\varepsilon_{\rm F})' \tau_2]'\tau_{\rm tr}/\varepsilon_{\rm F} \over 1+(\omega\tau_{\rm tr})^2}.
\end{align}
Here $\varepsilon_{\rm F}$ is the Fermi energy,  $\tau_{\rm tr}$ is the transport relaxation time which determines the electron mobility, $\tau_2$ is the relaxation time of the second angular harmonics of the electron distribution function, prime denotes differentiation over $\varepsilon_{\rm F}$, and  $\tau_{{\rm n}\omega} = \tau_{\rm n}/( 1-i\omega\tau_{\rm n})$ with ${\rm n}={\rm tr},2$. These expressions describe the frequency dependence of the  DCDR-induced ratchet current which varies for different disorder potentials~\cite{Nalitov2012}.

\section{Electrostatic simulations}
\label{App:D}

\begin{figure*}[t]
	\centering
	\includegraphics[width=\textwidth]{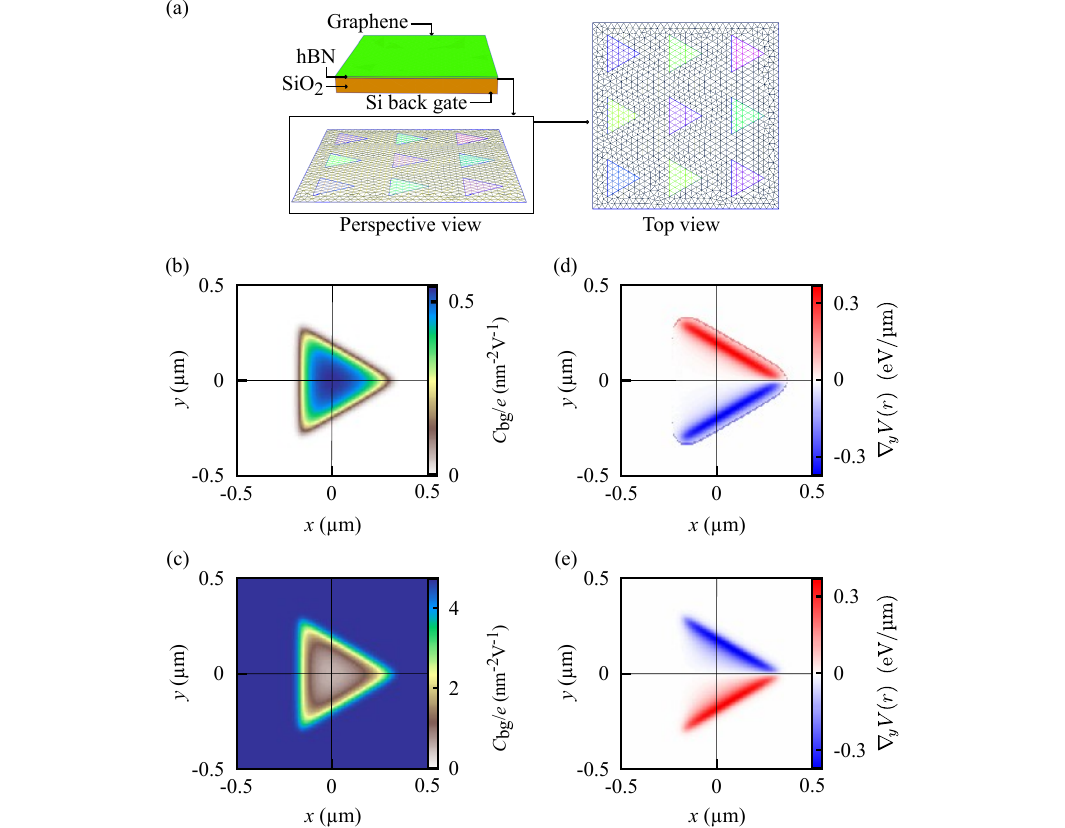}
	\caption{(a) Geometry of the modeled device for the electrostatically simulated and spatially dependent capacitance of (b) the global back gate and (c) the patterned  gate with only one period shown. Partial-differentiated onsite potential as a function of $x$ and $y$, considering (d) $(V_{\mathrm{bg}}, V_\mathrm{p}) = (1.5, 0)\ \mathrm{V}$ and (e) $(V_{\mathrm{bg}}, V_\mathrm{p}) = (0, 0.5)\ \mathrm{V}$.}
	\label{fig:electrostatics}
\end{figure*}

The modeled device for the electrostatic simulation is considered under the geometry of the experimental device, but it contains part of triangle pattern, that is a  $3 \times 3$ triangle array, as shown in the rectangular box in Fig.\,\ref{fig:electrostatics}(a). In addition, we assume that the regions in the holes are vacuums because the hBN and SiO$_{2}$ with the volume of such thin holes only slightly change the electric potential distribution. We set the dielectric constant of hBN and  SiO$_{2}$ to  3 and 3.9, respectively. Based on the periodicity of the triangle pattern, we focus on one period  and set the origin at the equilateral triangle center in the following analysis.

\textsc{FEniCS}~\cite{logg_automated_2012}, a Python package for the finite-element method, combined with the mesh generator \textsc{Gmsh}~\cite{geuzaine_gmsh_2009}, is adopted to numerically solve the Poisson equation in the modeled device and extract the electric displacement field near  graphene. Figures~\ref{fig:electrostatics}(b) and~(c) are the spatially dependent capacitance of the Si back gate and the graphite patterned  gate, respectively, calculated by dividing the induced carrier density $n(\bm r)$ by a unit gate voltage while the other gate is grounded. The induced carrier density can be calculated by the formula of surface charge density $\sigma(\bm{r})$ divided by the elementary charge
$$\sigma(\bm{r}) = \hat{\bm{n}}\cdot[\bm{D}_{\mathrm{out}}(\bm{r})-\bm{D}_{\mathrm{in}}(\bm{r})],$$
where $\hat{\bm{n}}$ is the unit normal vector of the surface of graphene and $\bm{D}$ is the electric displacement field.

With the gate capacitors simulated and stored, the gate-dependent onsite potential is obtained by
\begin{equation}
	V(\bm{r}) = -\mathrm{sgn}[n(\bm{r})]\hbar v_{0}\sqrt{\pi\vert n(\bm{r})\vert},
	\label{eqn:onsite_energy}
\end{equation}
where $v_{0}$ is the group velocity of the graphene, and the induced carrier density $n(\bm{r})$ is obtained by summing over the contributions from all gates without the need of bothering with the electrostatic simulation again. Figures~\ref{fig7} and\,\ref{fig:electrostatics}(d),\,(e) show the partial-differentiated onsite potential $\nabla_{x}V(\bm{r})$ or $\nabla_{y}V(\bm{r})$.

\bibliography{all_lib.bib}

\end{document}